\documentclass[10pt,a4paper]{article}

\usepackage[utf8]{inputenc}
\usepackage[english]{babel}

\usepackage{amsmath}
\usepackage{amsfonts}
\usepackage{amssymb}

\usepackage[colorlinks,citecolor=blue,urlcolor=blue,linkcolor=blue]{hyperref}

\usepackage[left=2cm,right=2cm,top=2cm,bottom=2cm]{geometry}

\usepackage{authblk}

\usepackage{feynmp}
\newcommand{\pd}{\partial}

\usepackage{graphicx}

\title{Effective potential of scalar-tensor gravity with quartic self-interaction of scalar field}
\author[1,2]{A. Arbuzov \thanks{arbuzov@theor.jinr.ru}}
\author[1,2]{B. Latosh \thanks{latosh@theor.jinr.ru}}
\author[1,2]{A. Nikitenko
\thanks{nikitenko@theor.jinr.ru}}
\affil[1]{Bogoliubov Laboratory of Theoretical Physics, JINR, Dubna, 141980, Russia}
\affil[2]{Dubna State University, Universitetskaya str. 19, Dubna, 141982, Russia}
\date{\today}

\begin{document}

\maketitle

\begin{abstract}
  One-loop effective potential of scalar-tensor gravity with a quartic scalar field self-interaction is evaluated up to first post-Minkowskian order. The potential develops an instability in the strong field regime which is expected from an effective theory. Depending on model parameters the instability region can be exponentially far in a strong field region. Possible applications of the model for inflationary scenarios are highlighted. It is shown that the model can enter the slow-roll regime with a certain set of parameters.
\end{abstract}

\section{Introduction}

Scalar-tensor models of gravity occupy a special place in the landscape of modified gravity. They provide a minimal modification of general relativity yet have a vast phenomenology. Valuable examples of scalar-tensor gravity are provided by inflation theories. A scalar field can drive inflation if it either has a potential of a special form or a non-minimal coupling with gravity~\cite{Starobinsky:1980te,Linde:1983gd,Kobayashi:2011nu,Starobinsky:2016kua,Gorbunov:2011zzc}.

Two following models should be highlighted due to their specific features. The first one is the Starobinsky inflation~\cite{Starobinsky:1980te}. This model extends general relativity with a curvature squared term $R^2$ which drives inflation. The term is generated at the one-loop level within quantum theory\footnote{At the one loop level the following operators are generated $R^2$, $R_{\mu\nu}^2$, $R_{\mu\nu\alpha\beta}^2$, $\square R$. The latter two shall be omitted as they are reduced to a complete derivative in four dimensions. $R_{\mu\nu}^2$ term can be brought to the Weyl tensor squared which is conformally invariant. Therefore it is irrelevant for Friedmann cosmologies as the Friedmann spacetime is conformally flat.} so it can be naturally included in a low-energy effective action. Thus one can argue that the Starobinsky model describes inflation as an effect dynamically generated at the loop level. Lastly, the Starobinsky inflation, together with any $f(R)$-gravity, can be mapped on a scalar-tensor 
gravity~\cite{Dicke:1961gz,Maeda:1988ab,Wands:1993uu,Magnano:1993bd,Faraoni:1998qx,DeFelice:2010aj,Calmet:2019odl}. 
The second model worth to be mentioned is the specific Horndeski model with the following kinetic coupling 
$G^{\mu\nu}\nabla_\mu\phi \nabla_\nu\phi$ where $G^{\mu\nu}$ is the Einstein tensor~\cite{Horndeski:1974wa,Horndeski:1976gi,Kobayashi:2011nu}. It has certain similarities with the Starobinsky inflation. Firstly, this kinetic coupling can drive inflation~\cite{Kobayashi:2011nu,Starobinsky:2016kua,Avdeev:2021von}. Secondly, such a non-minimal coupling is also dynamically induced at the one-loop level~\cite{Latosh:2020jyq}.

These models show that it is important to study quantum effects in scalar-tensor gravity as they may provide natural candidates for inflationary models. In particular, in paper~\cite{Arbuzov:2020pgp} one-loop effective potentials of some simplest scalar-tensor models were studied. It was found that they generate effective potentials which have areas with small slow-roll parameters therefore they may provide viable inflationary scenarios.

In this paper we study a more sophisticated scalar-tensor model and evaluate its effective potential. In the previous paper~\cite{Arbuzov:2020pgp} only models without scalar field self-interaction were considered for the sake of simplicity. 
Here we study the model given by the following action:
\begin{align}\label{the_model}
  \mathcal{A}=\int d^4 x \sqrt{-g} \,\left[-\cfrac{2}{\kappa^2}\, R -\cfrac12\, \phi\, (\square +m^2)\, \phi -\cfrac{\lambda}{4!}\,\phi^4\right] .
\end{align}
Here $\lambda$ is a dimensionless coupling constant chosen to be positive for the sake of stability; 
$\kappa$ is related to the Newton constant $G_N$:
\begin{align}
  \kappa^2 = 32 \pi G_N.
\end{align}
We also only consider the case with a non-broken symmetry $(m^2>0)$. The broken symmetry case requires a separate treatment and we will address this issue elsewhere.

There are several reasons to study this model. Firstly, the model serves as a natural extension of the previous studies \cite{Coleman:1973jx,Arbuzov:2020pgp,Latosh:2020jyq}. In particular, it would be possible to separate an influence of the scalar field self-interaction on an effective potential. Secondly, the model has only two dimensional parameters which are the Planck scale $\kappa^{-1}$ and the scalar field mass $m$. The scalar field self-interaction coupling $\lambda$ is dimensionless. This allows one to negate the issue of a mass scale hierarchy as it is natural to assume that the scalar field mass $m$ should be much smaller that the Planck mass. Thirdly, in a flat spacetime the scalar sector of model~\eqref{the_model} is renormalizable although gravity ruins this feature, as we will see. The effective potential technique provides a framework capable to soften some renormalization issues and to consistently study quantum effects associated with gravity. Finally, model~\eqref{the_model} can serve as an analogy to the Higgs sector of the Standard Model. It may provide a useful insight in Higgs inflation 
scenarios~\cite{Bezrukov:2007ep,Barvinsky:2008ia,Planck:2018jri,Bezrukov:2010jz}.

This paper is organized as follows. In the next section we briefly discuss the effective potential technique and apply it for model~\eqref{the_model}. We discuss the structure of the leading gravitational corrections and show that certain corrections can be resummed in an analogy to the well-known case~\cite{Coleman:1973jx}. In Section~\ref{section_discussion} we discuss features of the potential and its possible implementations for inflation. In Appendix~\ref{section_factors} we give a more detailed description of the calculation of the effective potential.

\section{Effective potential}

The effective potential technique is well-known within quantum field theory and very well presented in~\cite{Buchbinder:1992rb}. Because of this we will not discuss it in detail. However, a few comments on its implementation for gravity is due.

We constraint ourselves only to perturbative quantum gravity. Therefore we assume that the complete spacetime metric $g_{\mu\nu}$ is composed of a flat background $\eta_{\mu\nu}$ and small perturbations $h_{\mu\nu}$:
\begin{align}
  g_{\mu\nu} = \eta_{\mu\nu} + \kappa \, h_{\mu\nu}. 
\end{align}
The action describing gravity within perturbative approach shall be expanded in an infinite series with respect to $h_{\mu\nu}$. 
Therefore the complete action will contain an infinite number of terms describing $h_{\mu\nu}$ self-interactions and 
$h_{\mu\nu}$ couplings to matter. Not all of these terms are relevant in the effective potential context. 
Terms describing interactions of perturbations $h_{\mu\nu}$ can be omitted completely as they do not contribute to 
the effective potential of the scalar field at the one-loop level. Similarly, at the one-loop level among all terms 
describing matter gravitational coupling only terms linear in $\kappa$ are relevant. This allows one to reduce 
the infinite perturbative expansion of an action to an expression with a finite number of terms. Finally, the 
gravitational field gauge should be fixed. We use the following gauge-fixing term which allows one to decouple 
Faddeev-Popov ghosts:
\begin{align}
  \mathcal{A}_\text{gf} &=\int d^4 x \left(\pd_\mu h^{\mu\nu} - \cfrac12\, \pd^\mu h\right)^2 .
\end{align}

Therefore within the perturbative approach action \eqref{the_model} is reduced to the following form:
\begin{align}\label{the_model_expansion}
  \begin{split}
    \mathcal{A}+\mathcal{A}_\text{gf} =& \int d^4 x \Bigg[-\cfrac12\, h^{\mu\nu} \left(\cfrac12\, C_{\mu\nu\alpha\beta} \square \right) h^{\alpha\beta} - \cfrac12\, \phi\, (\square+m^2)\phi\\
      &- \cfrac{\lambda}{4!}\,\phi^4  -\cfrac{\kappa}{4} \, h^{\mu\nu} \left[ C_{\mu\nu\alpha\beta} \, \pd^\alpha\phi\,\pd^\beta\phi +\eta_{\mu\nu} \, m^2\, \phi^2 \right] + \cfrac{\kappa}{2} \, h^{\mu\nu} \, \eta_{\mu\nu} \cfrac{\lambda}{4!} \, \phi^4 \Bigg] + \text{irrelevant terms}.
  \end{split}
\end{align}
Here tensor $C$ is defined as follows:
\begin{align}
  C_{\mu\nu\alpha\beta} = \eta_{\mu\alpha}\eta_{\nu\beta}+\eta_{\mu\beta}\eta_{\nu\alpha}-\eta_{\mu\nu}\eta_{\alpha\beta}\,.
\end{align}
This action generates the following set of tree-level Feynman rules:
\begin{align}
  \begin{gathered}
    \begin{fmffile}{pic_A01}
      \begin{fmfgraph*}(30,30)
        \fmfleft{L}
        \fmfright{R}
        \fmf{dashes}{L,R}
      \end{fmfgraph*}
    \end{fmffile}
  \end{gathered}
  & = i\, \cfrac{1}{k^2-m^2}\,, 
  &
  \begin{gathered}
    \begin{fmffile}{pic_A02}
      \begin{fmfgraph*}(30,30)
        \fmfleft{L}
        \fmfright{R}
        \fmf{dbl_wiggly}{L,R}
        \fmflabel{$\mu\nu$}{L}
        \fmflabel{$\alpha\beta$}{R}
      \end{fmfgraph*}
    \end{fmffile}
  \end{gathered}
  &\hspace{.7cm} = i\,\cfrac{\, C_{\mu\nu\alpha\beta}}{k^2} \,, &
  \begin{gathered}
    \begin{fmffile}{pic_A03}
      \begin{fmfgraph*}(30,30)
        \fmfleft{L1,L2}
        \fmfright{R1,R2}
        \fmf{dashes}{L1,V,L2}
        \fmf{dashes}{R1,V,R2}
        \fmfdot{V}
      \end{fmfgraph*}
    \end{fmffile}
  \end{gathered}
  = -i \, \cfrac{\lambda}{4!}\, ,
\end{align}
\begin{align}
  \begin{gathered}
    \begin{fmffile}{pic_A04}
      \begin{fmfgraph*}(30,30)
        \fmfleft{L}
        \fmfright{R1,R2}
        \fmf{dbl_wiggly}{L,V}
        \fmf{dashes}{R1,V,R2}
        \fmflabel{$\mu\nu$}{L}
        \fmflabel{$p$}{R1}
        \fmflabel{$q$}{R2}
        \fmfdot{V}
      \end{fmfgraph*}
    \end{fmffile}
  \end{gathered}
  \hspace{.5cm}&=i\,\cfrac{\kappa}{4}\,\left(C_{\mu\nu\alpha\beta}\,p^\alpha\,q^\beta-\eta_{\mu\nu}\,m^2\right), &
  \begin{gathered}
    \begin{fmffile}{pic_A05}
      \begin{fmfgraph*}(40,40)
        \fmftop{T}
        \fmfleft{L1,L2}
        \fmfright{R1,R2}
        \fmf{dbl_wiggly,tension=.3}{T,V}
        \fmf{dashes}{L1,V,L2}
        \fmf{dashes}{R1,V,R2}
        \fmflabel{$\mu\nu$}{T}
        \fmfdot{V}
      \end{fmfgraph*}
    \end{fmffile}
  \end{gathered}
  = -i\,\cfrac{\kappa}{2}\,\eta_{\mu\nu}\, \cfrac{\lambda}{4!} \,.
\end{align}

We only consider the case with $m^2>0$ corresponding to a non-broken symmetry of the original action~\eqref{the_model}. There are both physical and technical considerations motivating such a decision. First and foremost, if the symmetry of the original action \eqref{the_model} is spontaneously broken and the scalar field potential has a well-known form $V(\phi) = \lambda \,\left(\phi^2 - \phi_0^2 \right)^2$, then one shall use different dynamical variables. The scalar field variable $\phi$ describes perturbations about a false vacuum which is unstable. Because of this the corresponding effective potential will be also unstable\footnote{In matter of fact, the evaluated expression for the effective potential~\eqref{the_effective_potential} becomes non-Hermitian if $m^2<0$.} which is a mere reflection of the poor choice of dynamical variables. Naturally, one shall use variables that describe scalar field perturbations about the true ground state which, in turn, will spawn a new set of Feynman rules that shall be used to evaluate the corresponding effective potential. Consequently, the broken symmetry case presents an independent problem which shall be addressed elsewhere.

From the physical point of view, the broken symmetry case also presents a separate problem. It is expected to find a broken symmetry in a low energy regime where a scalar field experience small perturbations about the true vacuum state. The non-broken symmetry case, on the other hand, is expected to describe a high energy regime when a scalar field experience strong excitations. The later case is more suitable for the inflation paradigm as at the beginning of inflation a scalar field is expected to be highly excited. Because of these reasons within this paper we only consider the non-broken symmetry case.

Let us turn to a discussion of the effective potential. To find the one-loop scalar field effective potential it is required, firstly, to find all $n$-point connected Green functions $G^c_n(x_1,\cdots,x_n)$; secondly, to obtain all $n$-point vertex functions $\Gamma_n(x_1,\cdots,x_n)$ from the corresponding Green functions via an amputation of external lines; thirdly, to calculate the effective action $\Gamma$ as the following series \cite{Buchbinder:1992rb}:
\begin{align}
  \Gamma = \sum\limits_{n=1}^\infty \, \cfrac{1}{n!} \, \int d^4 x_1 \cdots d^4 x_n\, \Gamma_n(x_1,\cdots,x_n) \, \varphi(x_1)\cdots\varphi(x_n).
\end{align}
Here $\varphi(x) = \langle 0 \lvert \widehat{\phi} \rvert 0 \rangle$ is the expectation value of the quantum scalar field $\phi$. This shows that an effective potential is generated by one-loop one-particle irreducible diagrams. In turn, a few topologically equivalent diagrams contribute to each connected Green function so the vertex functions receive an additional ``combinatorial'' factor that partially cancels $1/n!$ in each term of the effective action expansion.

In this section we will not discuss the calculation of these ``combinatorial'' factors but rather focus on the renormalization of the potential and its features. A more detailed discussion of the calculations is given in Appendix \ref{section_factors}.

The model \eqref{the_model} is well studied without gravity. We will refer to this case as to the leading order (LO) contribution as it is of the order $\mathcal{O}(\kappa^0)$. All corrections associated with quantum gravity appear at least at $\mathcal{O}(\kappa^2)$ level and will provide next-to-leading order (NLO) contribution. At the leading order the effective potential is generated by the following series:
\begin{align}
  \begin{gathered}
    \begin{fmffile}{pic_S1_01}
      \begin{fmfgraph*}(40,40)
        \fmfleft{L}
        \fmfright{R}
        \fmf{dashes}{L,V,R}
        \fmf{dashes,right=1}{V,V}
        \fmfdot{V}
      \end{fmfgraph*}
    \end{fmffile}
  \end{gathered}
  +
  \begin{gathered}
    \begin{fmffile}{pic_S1_02}
      \begin{fmfgraph*}(40,40)
        \fmfleft{L1,L2}
        \fmfright{R1,R2}
        \fmf{dashes}{L1,VL,L2}
        \fmf{dashes}{R1,VR,R2}
        \fmf{dashes,right=1,tension=.3}{VL,VR,VL}
        \fmfdot{VL,VR}
      \end{fmfgraph*}
    \end{fmffile}
  \end{gathered}
  +
  \begin{gathered}
    \begin{fmffile}{pic_S1_03}
      \begin{fmfgraph*}(40,40)
        \fmfleft{L1,L2,L3,L4}
        \fmfright{R1,R2,R3,R4}
        \fmftop{T1,T2,T3,T4}
        \fmfbottom{B1,B2,B3,B4}
        \fmf{dashes,tension=.4}{V1,V2,V3,V1}
        \fmf{dashes}{L2,V1,B2}
        \fmf{dashes}{B3,V2,R2}
        \fmf{dashes}{T2,V3,T3}
        \fmfdot{V1,V2,V3}
      \end{fmfgraph*}
    \end{fmffile}
  \end{gathered}
  +\cdots \to \Delta V_\text{eff,LO}. 
\end{align}
The expression for the effective potential reads
\begin{align}\label{CW_effective_potential}
  \Delta V_\text{eff,LO}=&  \cfrac12\,\int\cfrac{d^4_E k}{(2\pi)^4} \, \ln\left[1+\cfrac{\frac{\lambda}{2}\,\varphi^2}{k^2+m^2}\right] \\
  \begin{split}
    =&\, \cfrac{m^2}{2}\, \varphi^2\, \left(\cfrac{\lambda}{16\pi^2}\right) \left[ \cfrac{1}{d-4}-\cfrac34+\cfrac12\,\gamma-\cfrac32\,\ln 2\pi -\cfrac12\,\ln 2 +\cfrac12\ln\cfrac{\frac{\lambda}{2}\varphi^2 + m^2}{\mu^2}\right]\\
    & +\cfrac{\lambda}{4!} \, \varphi^4 \,\left(\cfrac{3\lambda}{16\pi^2}\right) \left[ \cfrac{1}{d-4}-\cfrac34+\cfrac12\,\gamma-\cfrac32\,\ln 2\pi -\cfrac12\,\ln 2 +\cfrac12\ln\cfrac{\frac{\lambda}{2}\varphi^2 + m^2}{\mu^2}\right]\\
    & +\cfrac{m^4}{64\pi^2} \, \ln\left[1+\cfrac{\frac{\lambda}{2}\,\varphi^2}{m^2}\right] + \mathcal{O}(d-4) .
  \end{split}
\end{align}
Within this paper we used dimensional regularization in $d$ dimensions with $\mu$ being the renormalization scale. Unlike the massless case studied in~\cite{Coleman:1973jx}, potential \eqref{CW_effective_potential} is regular in $\varphi\to 0$ limit due to the presence of a non-vanishing scalar field mass.

Next-to-leading order corrections to potential \eqref{CW_effective_potential} are controllable. Firstly, the one-loop effective potential is generated only by one-loop one-particle irreducible diagrams with external lines being all scalars carrying vanishing momenta. Because of this only a few terms from the infinite expansion of~\eqref{the_model} in a series with respect to $h_{\mu\nu}$ are relevant. Namely, terms describing self-interactions of $h_{\mu\nu}$ do not contribute to the potential. Secondly, among all terms describing gravitational coupling of the scalar field only terms linear in $\kappa$ are relevant. The other terms either contribute at higher orders of $\kappa$ or at the higher loop orders. Finally, because of the structure of the one-loop one-particle irreducible diagrams virtual gravitons (associated with $h_{\mu\nu}$) can only appear in the loop. These considerations allow one to split all possible one-loop one-particle irreducible diagrams in three separate series.
These three series contributing at next-to-leading order to the effective potential read
\begin{align}\label{series_1}
  \begin{gathered}
    \begin{fmffile}{pic_S2_01}
      \begin{fmfgraph}(50,50)
        \fmfleft{L}
        \fmfright{R}
        \fmf{dashes,tension=3}{L,V}
        \fmf{dashes,left=1}{V,VR}
        \fmf{dbl_wiggly,right=1}{V,VR}
        \fmf{dashes,tension=3}{VR,R}
        \fmfdot{V,VR}
      \end{fmfgraph}
    \end{fmffile}
  \end{gathered}
  +
  \begin{gathered}
    \begin{fmffile}{pic_S2_02}
      \begin{fmfgraph}(50,50)
        \fmfleft{L}
        \fmfright{R}
        \fmftop{T0,T1,T2,T3,T4}
        \fmfbottom{B}
        \fmf{dashes,tension=3}{L,V}
        \fmf{phantom}{T2,VT}
        \fmf{phantom,left=1}{V,VR}
        \fmf{dbl_wiggly,right=1}{V,VR}
        \fmf{dashes,tension=3}{VR,R}
        \fmf{phantom}{T2,VT}
        \fmf{phantom}{VT,B}
        \fmffreeze
        \fmf{dashes}{T1,VT,T3}
        \fmf{dashes}{V,VT,VR}
        \fmfdot{V,VR,VT}
      \end{fmfgraph}
    \end{fmffile}
  \end{gathered}
  +
  \begin{gathered}
    \begin{fmffile}{pic_S2_03}
      \begin{fmfgraph}(80,40)
        \fmfleft{L}
        \fmfright{R}
        \fmftop{T1,T2,T3,T4,T5,T6}
        \fmfbottom{B1,B2,B3,B4}
        \fmf{phantom,tension=3}{L,VL}
        \fmf{phantom,tension=3}{R,VR}
        \fmf{phantom}{VL,VR}
        \fmffreeze
        \fmf{phantom}{T2,VT1,T3}
        \fmf{phantom}{T4,VT2,T5}
        \fmf{phantom}{VT1,B2}
        \fmf{phantom}{VT2,B3}
        \fmf{phantom}{VT1,VT2}
        \fmffreeze
        \fmfdot{VL,VR,VT1,VT2}
        \fmf{dashes}{L,VL}
        \fmf{dashes}{R,VR}
        \fmf{dashes}{T2,VT1}
        \fmf{dashes}{T3,VT1}
        \fmf{dashes}{T4,VT2}
        \fmf{dashes}{T5,VT2}
        \fmf{dashes}{VL,VT1,VT2,VR}
        \fmf{dbl_wiggly,right=.4}{VL,VR}
      \end{fmfgraph}
    \end{fmffile}
  \end{gathered}
  +\cdots
  \to\Delta V_\text{eff,NLO,1} \,,
\end{align}
\begin{align}\label{series_2}
  \begin{gathered}
    \begin{fmffile}{pic_S3_01}
      \begin{fmfgraph}(50,50)
        \fmfleft{L0,L1,L2,L3,L4}
        \fmfright{R}
        \fmf{dashes}{L1,V}
        \fmf{dashes}{L2,V}
        \fmf{dashes}{L3,V}
        \fmf{dashes,left=1}{V,VR}
        \fmf{dbl_wiggly,right=1}{V,VR}
        \fmf{dashes,tension=2}{VR,R}
        \fmfdot{V,VR}
      \end{fmfgraph}
    \end{fmffile}
  \end{gathered}
  +
  \begin{gathered}
    \begin{fmffile}{pic_S3_02}
      \begin{fmfgraph}(50,50)
        \fmfleft{L0,L1,L2,L3,L4}
        \fmfright{R}
        \fmftop{T0,T1,T2,T3,T4}
        \fmfbottom{B}
        \fmf{dashes}{L1,V}
        \fmf{dashes}{L2,V}
        \fmf{dashes}{L3,V}
        \fmf{phantom}{T2,VT}
        \fmf{phantom,left=1}{V,VR}
        \fmf{dbl_wiggly,right=1}{V,VR}
        \fmf{dashes,tension=2.7}{VR,R}
        \fmf{phantom}{T2,VT}
        \fmf{phantom}{VT,B}
        \fmffreeze
        \fmf{dashes}{T1,VT,T3}
        \fmf{dashes}{V,VT,VR}
        \fmfdot{V,VR,VT}
      \end{fmfgraph}
    \end{fmffile}
  \end{gathered}
  +
  \begin{gathered}
    \begin{fmffile}{pic_S3_03}
      \begin{fmfgraph}(80,40)
        \fmfleft{L0,L1,L2,L3,L4}
        \fmfright{R}
        \fmftop{T1,T2,T3,T4,T5,T6}
        \fmfbottom{B1,B2,B3,B4}
        \fmf{phantom,tension=3}{L2,VL}
        \fmf{phantom,tension=3}{VR,R}
        \fmf{phantom}{VL,VR}
        \fmf{dashes}{T2,VT1}
        \fmf{dashes}{T3,VT1}
        \fmf{dashes}{T4,VT2}
        \fmf{dashes}{T5,VT2}
        \fmf{phantom}{VT1,B2}
        \fmf{phantom}{VT2,B3}
        \fmf{phantom}{VT1,VT2}
        \fmffreeze
        \fmf{dashes}{L1,VL}
        \fmf{dashes}{L2,VL}
        \fmf{dashes}{L3,VL}
        \fmf{dashes}{VR,R}
        \fmf{dbl_wiggly,right=.5}{VL,VR}
        \fmfdot{VL,VR,VT1,VT2}
        \fmf{dashes}{VL,VT1,VT2,VR}
      \end{fmfgraph}
    \end{fmffile}
  \end{gathered}
  +\cdots
  \to\Delta V_\text{eff,NLO,2} \,.
\end{align}
\begin{align}\label{series_3}
  \begin{gathered}
    \begin{fmffile}{pic_S4_01}
      \begin{fmfgraph}(50,50)
        \fmfleft{L0,L1,L2,L3,L4}
        \fmfright{R0,R1,R2,R3,R4}
        \fmf{dashes}{L1,V}
        \fmf{dashes}{L2,V}
        \fmf{dashes}{L3,V}
        \fmf{dashes,left=1}{V,VR}
        \fmf{dbl_wiggly,right=1}{V,VR}
        \fmf{dashes}{VR,R1}
        \fmf{dashes}{VR,R2}
        \fmf{dashes}{VR,R3}
        \fmfdot{V,VR}
      \end{fmfgraph}
    \end{fmffile}
  \end{gathered}
  +
  \begin{gathered}
    \begin{fmffile}{pic_S4_02}
      \begin{fmfgraph}(50,50)
        \fmfleft{L0,L1,L2,L3,L4}
        \fmfright{R0,R1,R2,R3,R4}
        \fmftop{T0,T1,T2,T3,T4}
        \fmfbottom{B}
        \fmf{dashes}{L1,V}
        \fmf{dashes}{L2,V}
        \fmf{dashes}{L3,V}
        \fmf{phantom}{T2,VT}
        \fmf{phantom,left=1}{V,VR}
        \fmf{dbl_wiggly,right=1}{V,VR}
        \fmf{dashes}{VR,R1}
        \fmf{dashes}{VR,R2}
        \fmf{dashes}{VR,R3}
        \fmf{phantom}{T2,VT}
        \fmf{phantom}{VT,B}
        \fmffreeze
        \fmf{dashes}{T1,VT,T3}
        \fmf{dashes}{V,VT,VR}
        \fmfdot{V,VR,VT}
      \end{fmfgraph}
    \end{fmffile}
  \end{gathered}
  +
  \begin{gathered}
    \begin{fmffile}{pic_S4_03}
      \begin{fmfgraph}(80,40)
        \fmfleft{L0,L1,L2,L3,L4}
        \fmfright{R0,R1,R2,R3,R4}
        \fmftop{T1,T2,T3,T4,T5,T6}
        \fmfbottom{B1,B2,B3,B4}
        \fmf{phantom,tension=3}{L2,VL}
        \fmf{phantom,tension=3}{VR,R2}
        \fmf{phantom}{VL,VR}
        \fmf{dashes}{T2,VT1}
        \fmf{dashes}{T3,VT1}
        \fmf{dashes}{T4,VT2}
        \fmf{dashes}{T5,VT2}
        \fmf{phantom}{VT1,B2}
        \fmf{phantom}{VT2,B3}
        \fmf{phantom}{VT1,VT2}
        \fmffreeze
        \fmf{dashes}{L1,VL}
        \fmf{dashes}{L2,VL}
        \fmf{dashes}{L3,VL}
        \fmf{dashes}{VR,R1}
        \fmf{dashes}{VR,R2}
        \fmf{dashes}{VR,R3}
        \fmf{dbl_wiggly,right=.5}{VL,VR}
        \fmfdot{VL,VR,VT1,VT2}
        \fmf{dashes}{VL,VT1,VT2,VR}
      \end{fmfgraph}
    \end{fmffile}
  \end{gathered}
  +\cdots
  \to\Delta V_\text{eff,NLO,3} \,.
\end{align}
In series \eqref{series_1} we collect all contributions which are spawned by the gravitational interaction with the scalar field kinetic energy, described by term $\sqrt{-g}\,g^{\mu\nu} \nabla_\mu\phi\nabla_\nu\phi$. To be exact, the first term of series~\eqref{series_1} is generated only by the gravitational scalar self-energy which is due to its kinetic energy. The other terms of the series only introduce contributions related with the scalar field self-interaction. In series~\eqref{series_2} we collect terms that are generated both by kinetic and potential energy. The potential energy, in turn, is described by term $\sqrt{-g} \phi^4$. Finally, in series~\eqref{series_3} we collect contributions that are generated by pure potential energy. The corresponding contributions to the effective potential read
\begin{align}
  \begin{split}
    \Delta V_\text{eff,NLO,1} =& \cfrac{\kappa^2 m^4}{\lambda} \, \int\cfrac{d^4_E k}{(2\pi)^4}\,\cfrac{1}{k^2} \, \cfrac{\cfrac{\frac{\lambda}{2} \varphi^2}{k^2+m^2}}{1+\cfrac{\frac{\lambda}{2} \varphi^2}{k^2+m^2}} \\
    =& \cfrac{m^2}{2} \, \varphi^2 \, \cfrac{\kappa^2 m^2}{8\pi^2} \,\left[-\cfrac{1}{d-4}+\cfrac12\,-\cfrac12\,\gamma+\cfrac32\,\ln2\pi + \cfrac12\,\ln 2-\cfrac12\,\ln\cfrac{\frac{\lambda}{2}\,\varphi^2+m^2}{\mu^2}\right] + \mathcal{O}(d-4)\,,
  \end{split}\\
  \begin{split}
    \Delta V_\text{eff,NLO,2}=& \cfrac{\kappa^2 m^2}{3} \, \varphi^2 \int\cfrac{d^4_E k}{(2\pi)^4} \,\cfrac{1}{k^2} \, \cfrac{\cfrac{\frac{\lambda}{2} \, \varphi^2}{k^2+m^2}}{1+\cfrac{\frac{\lambda}{2} \, \varphi^2}{k^2+m^2}}\\
    =& \cfrac{\lambda}{4!}\,\varphi^4  \, \cfrac{\kappa^2 m^2}{2\pi^2}\left[-\cfrac{1}{d-4} +\cfrac12\,-\cfrac12\,\gamma+\cfrac32\,\ln2\pi + \cfrac12\,\ln 2-\cfrac12\,\ln\cfrac{\frac{\lambda}{2}\,\varphi^2+m^2}{\mu^2}\right] + \mathcal{O}(d-4),
  \end{split}\\
  \begin{split}
    \Delta V_\text{eff,NLO,3}=&\cfrac{\kappa^2\,\lambda}{36} \, \varphi^4\int\cfrac{d^4_E k}{(2\pi)^4} \,\cfrac{1}{k^2} \, \cfrac{\cfrac{\frac{\lambda}{2} \, \varphi^2}{k^2+m^2}}{1+\cfrac{\frac{\lambda}{2} \, \varphi^2}{k^2+m^2}}\\
    =&\cfrac{\kappa^2}{6!} \, \varphi^6 \, \cfrac{5 \lambda^2}{4\pi^2}\,\left[-\cfrac{1}{d-4} +\cfrac12\,-\cfrac12\,\gamma+\cfrac32\,\ln2\pi + \cfrac12\,\ln 2-\cfrac12\,\ln\cfrac{\frac{\lambda}{2}\,\varphi^2+m^2}{\mu^2}\right] + \mathcal{O}(d-4).
  \end{split}
\end{align}
It can be seen that these contributions share the expression in square brackets, which is not a coincidence. From the form of the corresponding diagrams~\eqref{series_1}, \eqref{series_2}, and~\eqref{series_3} it can be clearly seen that these series share the divergent structure.

The 1-loop contribution to the effective potential of model \eqref{the_model} evaluated up to $\mathcal{O}(\kappa^2)$ order reads
\begin{align}
  \Delta V_\text{eff} =&\cfrac12\,\int\cfrac{d^4_E k}{(2\pi)^4} \, \ln\left[1+\cfrac{\frac{\lambda}{2}\,\varphi^2}{k^2+m^2}\right] + \cfrac{\kappa^2 m^2}{\lambda} \left(m^2 + \cfrac{\lambda}{3} \,\varphi^2+\cfrac{\lambda^2\,\varphi^4}{36\,m^2}\right)\,\int\cfrac{d^4_E k}{(2\pi)^4} \, \cfrac{1}{k^2} \, \cfrac{\cfrac{\frac{\lambda}{2} \, \varphi^2}{k^2+m^2}}{1+\cfrac{\frac{\lambda}{2} \, \varphi^2}{k^2+m^2}} + \mathcal{O}(\kappa^4)\\
  \begin{split}\label{the_effective_potential}
    =&\cfrac{m^4}{64\pi^2} \ln\left[1+\cfrac{\frac{\lambda}{2}\,\varphi^2}{m^2}\right] + \cfrac{m^2}{2}\,\varphi^2 \left[\left(\cfrac{\lambda}{16\pi^2} - \cfrac{\kappa^2 m^2}{8\pi^2}\right)\left(\cfrac{1}{d-4} +\cfrac12\,\ln\cfrac{\frac{\lambda}{2}\,\varphi^2 + m^2}{\mu^2}\right)+\mathcal{F}_1\right]\\
    & +\cfrac{\lambda}{4!} \, \varphi^4 \left[\left(\cfrac{3\lambda}{16\pi^2} -\cfrac{\kappa^2 m^2}{2\pi^2}\right)\left(\cfrac{1}{d-4} +\cfrac12\,\ln\cfrac{\frac{\lambda}{2}\,\varphi^2 + m^2}{\mu^2}\right)+\mathcal{F}_2\right] \\
    & +\cfrac{\kappa^2}{6!} \, \varphi^6 \left[-\left(\cfrac{5 \lambda^2}{4\pi^2}\right)\left(\cfrac{1}{d-4} + \cfrac12\,\ln\cfrac{\frac{\lambda}{2}\,\varphi^2 + m^2}{\mu^2}\right) +\mathcal{F}_3\right] +\mathcal{O}(\kappa^4) 
  \end{split}
\end{align}
with $\mathcal{F}_1$, $\mathcal{F}_2$, and $\mathcal{F}_3$ being finite parts defined as follows:
\begin{align}
  \begin{split}
    \mathcal{F}_1 =& \cfrac{\lambda}{16\pi^2} \left\{-\cfrac34\,+\cfrac12\,\gamma-\cfrac32\,\ln(2\pi) -\cfrac12\,\ln(2)\right\}+\cfrac{\kappa^2 m^2}{8\pi^2} \left\{\cfrac12-\cfrac12\,\gamma+\cfrac32\ln(2\pi)+\cfrac12\,\ln(2)\right\} \, ,\\
    \mathcal{F}_2 =& \cfrac{3\,\lambda}{16\pi^2} \left\{-\cfrac34\,+\cfrac12\,\gamma-\cfrac32\,\ln(2\pi) -\cfrac12\,\ln(2)\right\}+\cfrac{\kappa^2 m^2}{2\pi^2} \left\{\cfrac12-\cfrac12\,\gamma+\cfrac32\ln(2\pi)+\cfrac12\,\ln(2)\right\} \,,\\
    \mathcal{F}_3 =& \cfrac{5 \lambda^2}{4\pi^2}\, \left\{ \cfrac12\,-\cfrac12\,\gamma+\cfrac32\,\ln2\pi + \cfrac12\,\ln 2\right\} \, .
  \end{split}
\end{align}

Expression \eqref{the_effective_potential} has a finite number of divergences, namely, only $\varphi^2$, $\varphi^4$, and $\varphi^6$ terms shall be renormalized. Unlike the flat spacetime case~\cite{Coleman:1973jx}, new interaction $\varphi^6$ is generated and it should be renormalized. The corresponding renormalization problems are softened within the effective theory approach. The generation of new non-renormalizable interactions at the loop level is expected as the theory contains quantum gravity which by itself is non-renormalizable. However, the effective potential (and the effective action) drives macroscopic degrees of freedom $\varphi=\langle 0 \lvert \phi \rvert 0 \rangle$ and it is not renormalized in the same sense as the microscopic action does. One shall choose a normalization point $\varphi_0$ and renormalize all divergent interactions on the values empirically measured in this point. But this procedure do not impose any constraints on fundamental physics, it only imposes the initial conditions taken from experiment.

We choose $\varphi_0=0$ as the normalization point. It can be done because of the presence of a non-vanishing scalar field mass. Let us note that with such a choice of the normalization point one shall pay a special attention to $m\to 0$ limit. In that limit $\varphi_0=0$ can no longer be a suitable normalization point, so the whole effective potential can become singular. Further, such a choice of a normalization point corresponds to a physical setup where the scalar field mass and self-interaction couplings are measured in the low-energy regime. We normalize the effective potential on the observed scalar field mass $m$, observed four-particle coupling $\lambda$, and observed six-particle coupling $g$ with an introduction of the following counter-term:
\begin{align}\label{the_counter-term}
  \begin{split}
    \Delta V_\text{c.t.} =&  \cfrac{g}{6!}\,\varphi^6-\cfrac{m^2}{2}\,\varphi^2 \left[\left(\cfrac{\lambda}{16\pi^2} - \cfrac{\kappa^2 m^2}{8\pi^2}\right)\left(\cfrac{1}{d-4} +\cfrac12\,\ln\cfrac{m^2}{\mu^2}\right)+\mathcal{F}_1+\cfrac{\lambda}{64\pi^2}\right]\\
    & -\cfrac{\lambda}{4!} \, \varphi^4 \left[\left(\cfrac{3\lambda}{16\pi^2} -\cfrac{\kappa^2 m^2}{2\pi^2}\right)\left(\cfrac{1}{d-4} +\cfrac12\,\ln\cfrac{m^2}{\mu^2}\right)+\mathcal{F}_2+\cfrac{3}{64\pi^2} \left(3\lambda-8\,\kappa^2 m^2 \right)\right]\\
    & -\cfrac{\kappa^2}{6!} \, \varphi^6 \left[ -\left(\cfrac{5\lambda^2}{4\pi^2}\right)\left(\cfrac{1}{d-4} + \cfrac12 \, \ln\cfrac{m^2}{\mu^2}\right) + \mathcal{F}_3\right] \, .
  \end{split}
\end{align}
The renormalized effective potential reads
%%
%\begin{align}\label{the_effective_potential_renormalized}
%  \begin{split}
%    V_\text{eff,ren} = \cfrac{m_0^2}{2}\, \varphi^2 + \cfrac{\lambda_0}{4!}\, \varphi^4 + \cfrac{g_0}{6!} \, \varphi^6-\cfrac{m^2}{2}\,\varphi^2\,\cfrac{\lambda}{64\pi^2} - \cfrac{\lambda}{4!} \, \varphi^4 \,\cfrac{3}{64\pi^2}\, (3\lambda - 8 \kappa^2 m^2) -\cfrac{\varphi^6}{6!}\,\cfrac{15\lambda}{32 \pi^2} \,\left(\cfrac{\lambda^2}{m^2} - 2 \kappa^2 \lambda \right) \\
%    + \ln\left[1+\cfrac{\frac{\lambda}{2}\,\varphi^2}{m^2} \right]\! \left\{\cfrac{m^4}{64\pi^2} + \cfrac{m^2}{2}\,\varphi^2 \, \cfrac12\left(\cfrac{\lambda}{16\pi^2} - \cfrac{\kappa^2 m^2}{8\pi^2}\right) + \cfrac{\lambda}{4!} \,\varphi^4\, \cfrac12\left(\cfrac{3\lambda}{16\pi^2}-\cfrac{\kappa^2 m^2}{2\pi^2}\right)-\cfrac{\kappa^2}{6!}\,\varphi^6\,\cfrac12\left(\cfrac{5\lambda^2}{4\pi^2}\right) \right\}
%  \end{split}
%\end{align}
%\begin{align}
%  =& \cfrac{m_0^2}{2}\, \varphi^2 +\cfrac{\lambda_0}{4!}\, \varphi^4 +\cfrac{g_0}{6!}\,\varphi^6 -\cfrac{\lambda ^3 \left(3 \lambda +16 \kappa ^2 m^2\right)}{36864 \pi ^2 m^4} \, \varphi^8 + \mathcal{O}\left(\varphi^{10}\right).
%\end{align}
%%
\begin{align}\label{the_effective_potential_renormalized_minimally}
  \begin{split}
    V_\text{eff,ren}=&\ln\left[1+\cfrac{\frac{\lambda}{2}\,\varphi^2}{m^2} \right] \left\{\cfrac{m^4}{64\pi^2} + \cfrac{m^2}{2}\,\varphi^2\,\cfrac{\lambda-2 m^2 \kappa^2}{32\pi^2} + \cfrac{\lambda}{4!} \,\varphi^4\, \cfrac{3\lambda - 8 m^2 \kappa^2}{32\pi^2}-\cfrac{\kappa^2}{6!} \, \varphi^6 \, \cfrac{5 \lambda^2}{8\pi^2} \right\}\\
    &+\cfrac{m^2}{2}\,\left(1- \cfrac{\lambda}{64\pi^2}\right)\, \varphi^2 + \cfrac{\lambda}{4!}\,\left(1 - \cfrac{3(3\lambda - 8 \kappa^2 m^2)}{64\pi^2} \right)\, \varphi^4 +\cfrac{g}{6!}\,\varphi^6\left(1-\cfrac{15\,\lambda^2}{32\pi^2}\,\cfrac{\lambda-2 m^2 \kappa^2}{g\,m^2}\right)
  \end{split}\\
  =& \cfrac{m^2}{2}\, \varphi^2 +\cfrac{\lambda}{4!}\, \varphi^4 +\cfrac{g}{6!}\,\varphi^6 -\cfrac{\lambda ^3 \left(3 \lambda +16 \kappa ^2 m^2\right)}{36864 \pi ^2 m^4} \, \varphi^8 + \mathcal{O}\left(\varphi^{10}\right).
\end{align}

The following comments on the renormalization shall be given. Firstly, we shall note that in the most general case one can distinguish the microscopic action mass $m$ and coupling $\lambda$ from the effective action (effective potential) mass $m_0$ and couplings $\lambda_0$, $g_0$. We believe that such an approach shall not be used in our case as it is physically irrelevant. The symmetry of the original model is not broken therefore both $\phi$ and $\varphi = \langle 0 \lvert \widehat{\phi} \rvert 0 \rangle$ describe scalar field excitations about the same state. Therefore there is no reason to believe that the effective action will develop parameters different from $m$ and $\lambda$. The new scalar field interaction $\varphi^6$, on the other hand, is generated dynamically by the effective theory. Because of this we can only normalize the corresponding coupling to the value $g$ that supposed to be measured in the low energy limit. Therefore the renormalized effective potential has only four free parameters $m$, $\lambda$, $g$, and $\kappa$ which are measure in normalization point $\varphi_0=0$.

Secondly, we shall draw attention to $m\to 0$ limit. As it was noted above, $\varphi_0=0$ is a suitable renormalization point if and only if the scalar field has a non-vanishing mass $m\not =0$. Therefore, if one takes $m\to 0$ limit they should choose a different normalization point and introduce a different counter-term. It can be clearly seen that before the renormalization effective 
potential~\eqref{the_effective_potential} remains regular in $m\to 0$ limit while the counter-term~\eqref{the_counter-term} is not. Therefore the absence of a direct $m\to 0$ limit in~\eqref{the_effective_potential_renormalized_minimally} is due to the choice of the normalization point only.

To proceed with the study we shall note that expression~\eqref{the_effective_potential_renormalized_minimally} contains four free parameters which are the scalar field mass $m$, scalar field quartic coupling $\lambda$, gravitational coupling $\kappa$, and scalar field dimensional coupling $g$. It is more convenient to analyze this expression in the scalar field mass units:
\begin{align}\label{the_effective_potential_renormalized_minimally_reduced_variables}
  \begin{split}
    \widetilde{V}_\text{eff,ren}=&\ln\left[1+\cfrac{\lambda}{2}\,\tilde{\varphi}^2 \right] \left\{\cfrac{1}{64\pi^2} + \cfrac{1}{2}\,\tilde{\varphi}^2\,\cfrac{\lambda-2 \zeta^2}{32\pi^2} + \cfrac{\lambda}{4!} \,\tilde\varphi^4\, \cfrac{3\lambda - 8\, \zeta^2}{32\pi^2}-\cfrac{1}{6!} \, \tilde\varphi^6 \, \cfrac{5 \lambda^2\,\zeta^2}{8\pi^2} \right\}\\
    &+\cfrac{1}{2}\,\left(1- \cfrac{\lambda}{64\pi^2}\right)\, \tilde\varphi^2 + \cfrac{\lambda}{4!}\,\left(1 - \cfrac{3(3\lambda - 8\,\zeta^2)}{64\pi^2} \right)\, \tilde\varphi^4 +\cfrac{\tilde{g}}{6!}\,\tilde\varphi^6\left(1-\cfrac{15\,\lambda^2}{32\pi^2}\,\cfrac{\lambda-2 \zeta^2}{\tilde{g}}\right)\,.
  \end{split}
\end{align}
Here $V_\text{eff,ren}=m^4\, \widetilde{V}_\text{eff,ren}$, $\varphi = m\, \widetilde{\varphi}$, $\zeta = \kappa \,m$, $\tilde{g}=g\, m^2$. We naturally assume that $\zeta \ll 1$ and the scalar field mass is much smaller that the Planck scale.

First and foremost, let us address the asymptotic of effective potential \eqref{the_effective_potential_renormalized_minimally_reduced_variables}. In the weak field region $\tilde\varphi\sim 0$, 
the potential is regular due to the presence of a non-vanishing mass:
\begin{align}
  \widetilde{V}_\text{eff,ren} =\cfrac12\,\tilde\varphi^2 + \cfrac{\lambda}{4!} \, \tilde\varphi^4+\cfrac{\tilde{g}}{6!} \, \tilde\varphi^6 +\mathcal{O}(\tilde\varphi^8)\,.
\end{align}
The first two terms of the expansion always have positive coefficients while $\tilde{g}$ may be negative. This shows that the potential may have a local maximum and an instability region. The existence of an instability region do not pose a serious problem within an effective theory as the region can lie outside the theory applicability domain.

The effective potential does develop an instability in the strong field region ($\tilde\varphi\to\infty$). 
In the strong field regime the leading contribution to the effective potential reads
\begin{align}\label{strong_field_asymptotic}
  \widetilde{V}_\text{eff,ren} \overset{\tilde\varphi\sim\infty}{\sim}\cfrac{1}{6!}\,\tilde\varphi^6 \left\{\tilde{g}\left(1-\cfrac{15\lambda^2}{32\pi^2}\,\cfrac{\lambda-2\,\zeta^2}{\tilde{g}}\right)-\cfrac{5 \lambda^2 \,\zeta^2}{8\pi^2} \ln\left[1+\cfrac{\lambda}{2}\, \tilde\varphi^2\right]\right\}\,.
\end{align}
The $\log$-term coefficient of this expression is always negative thus the potential always approaches $-\infty $ in $\tilde\varphi\to\infty$ limit. At the same time the first term in the brackets can be both negative and positive. If the term is negative, then~\eqref{strong_field_asymptotic} is negative even in the weak field regime. Consequently, the potential develops a local maximum that marks the stable region in a vicinity of $\tilde\varphi=0$. On the contrary, if the term is positive, then \eqref{strong_field_asymptotic} is positive in the weak field region and becomes negative only at $\tilde\varphi \gg 1$. The potential, in tern, develops a local maximum far in the strong field region.

The field value $\tilde\varphi_\text{crit}$ where \eqref{strong_field_asymptotic} becomes negative we call the critical value. The value can be derived from \eqref{strong_field_asymptotic}:
\begin{align}\label{critical_field_equation}
  \left(\tilde\varphi_\text{crit}\right)^2 = \cfrac{2}{\lambda}\left[\exp\left[\cfrac{8\pi^2\,\tilde{g}}{5\,\lambda^2\zeta^2}\left(1-\cfrac{15\lambda^2}{32\pi^2}\,\cfrac{\lambda-2\,\zeta^2}{\tilde{g}}\right)\right]-1\right]\,.
\end{align}
The critical field value is real if the following condition is met
\begin{align}\label{criticality_condition_0}
  \cfrac{8\pi^2\,\tilde{g}}{5\,\lambda^2\zeta^2}\left(1-\cfrac{15\lambda^2}{32\pi^2}\,\cfrac{\lambda-2\,\zeta^2}{\tilde{g}}\right)<0\,.
\end{align}
This condition is satisfied if either of the following two conditions is fulfilled:
\begin{align}
  0<\tilde{g}<\cfrac{15\lambda^2}{32\pi^2}\,(\lambda-2\zeta^2)\, , \label{criticality_condition_1} \\
  \tilde{g}<0 \text{ and } \tilde{g}<\cfrac{15\lambda^2}{32\pi^2}\,(\lambda-2\zeta^2)\,. \label{criticality_condition_2}
\end{align}

These results show the following behavior of the effective potential. In the weak field region the effective potential growth with $\tilde\varphi$. It growth to the local maximum point $\tilde\varphi_\text{max}$ and then strives to $-\infty$. The position of the local maximum is related with the strong field asymptotic. If either~\eqref{criticality_condition_1} or~\eqref{criticality_condition_2} is met, then the local maximum $\tilde\varphi_\text{max}$ exists in the weak field region. This is due to the fact that 
contribution~\eqref{strong_field_asymptotic} is negative even at $\tilde\varphi\sim 0$. 
If conditions~\eqref{criticality_condition_1} and~\eqref{criticality_condition_2} are not hold, then the local maximum 
is expected to appear about $\tilde\varphi_\text{crit}$~\eqref{critical_field_equation} which is exponential in model parameters and resides far in the strong field region.

\begin{figure}[htb]
  \begin{center}
    \includegraphics[width=\textwidth]{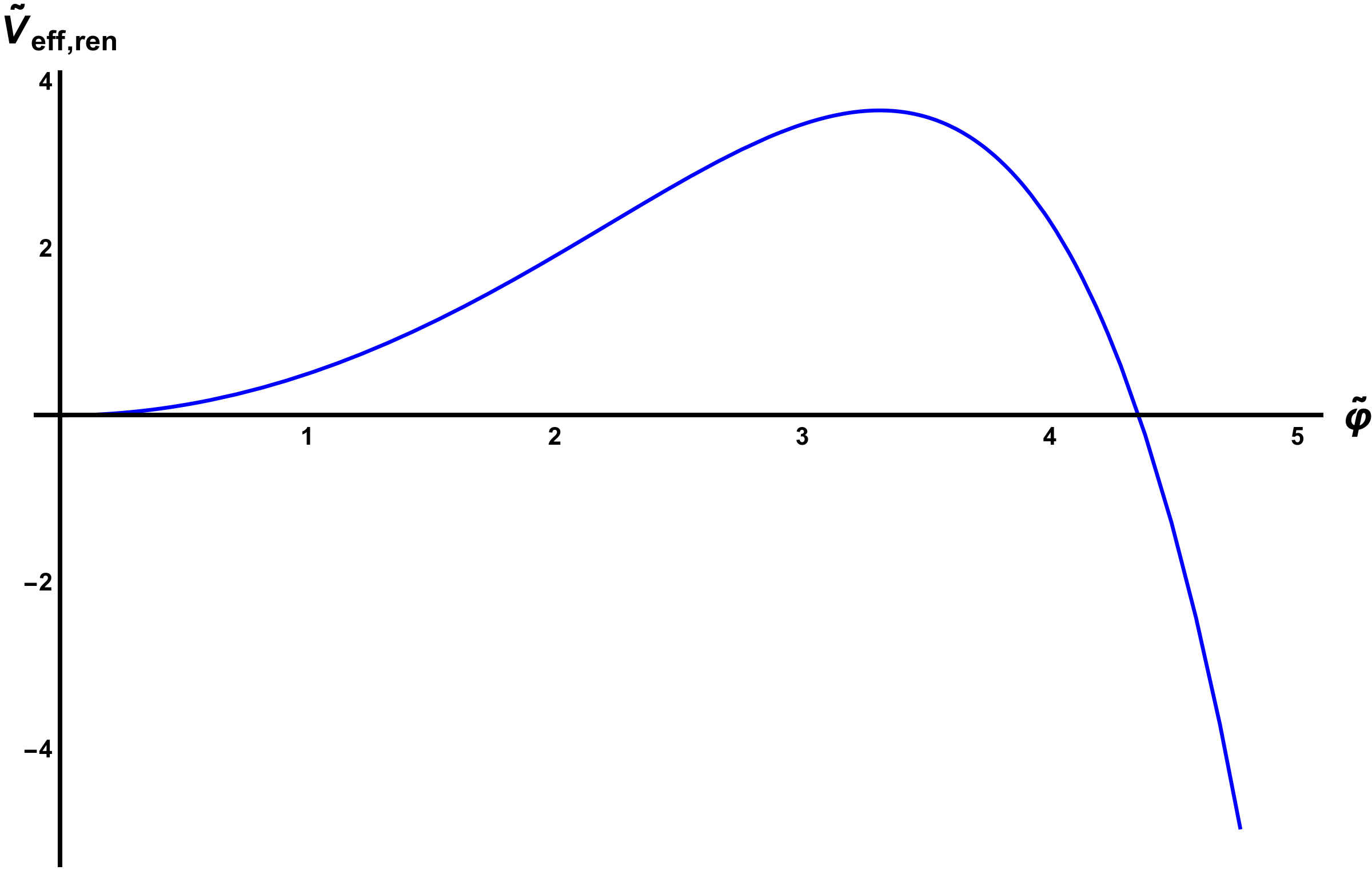}
  \end{center}
  \caption{Effective potential \eqref{the_effective_potential_renormalized_minimally_reduced_variables} with $\lambda=10^{-3}$, $\zeta=10^{-6}$, $\tilde{g}=-1$. The local maximum is $\tilde\varphi_\text{max} \simeq 3.31126$.}\label{plot_1}
\end{figure}

Because of the complicated form of potential \eqref{the_effective_potential_renormalized_minimally_reduced_variables}, 
it is more suitable to study it numerically. We present an explicit form of the effective 
potential~\eqref{the_effective_potential_renormalized_minimally_reduced_variables} with $\lambda=10^{-3}$, $\zeta=10^{-6}$, 
and $\tilde{g}=-1$ in Figure~\ref{plot_1}. With such parameters condition~\eqref{criticality_condition_2} is fulfilled 
and the local maximum appears in the weak field region $\tilde\varphi_\text{max} \simeq 3.31126$.

We would like to demonstrate a plot showing the exponential behavior of the local maximum but this is impossible as $\tilde\varphi_\text{crit}$ grows extremely rapidly. Let us take parameter $\lambda=10^{-3}$, $\zeta=10^{-6}$ from the previous case for the sake of illustration. We shall consider $\tilde{g}>0$ case which does not satisfy~\eqref{criticality_condition_1}. 
Condition~\eqref{criticality_condition_1} puts the following constraint on $\tilde{g}$:
\begin{align}
  \tilde{g} < \left. \cfrac{15\lambda^2}{32\pi^2}\,(\lambda-2\zeta^2) \right|_{\lambda=10^{-3},\, \zeta=10^{-6}} \sim 4.74 \times 10^{-11}\,.
\end{align}
If we set $\tilde{g} = 10^{-10}$, which is by a single order of amplitude bigger then the critical value, then the critical field value $\tilde\varphi_\text{crit}$ will be separated from the weak field region by many orders of magnitude:
\begin{align}
   \tilde\varphi_\text{crit} \Big|_{\lambda=10^{-3},\, \zeta=10^{-6}} \sim 2.8 \times 10^{360089499}\,.
\end{align}

\begin{figure}[htb]
  \begin{center}
    \includegraphics[width=\textwidth]{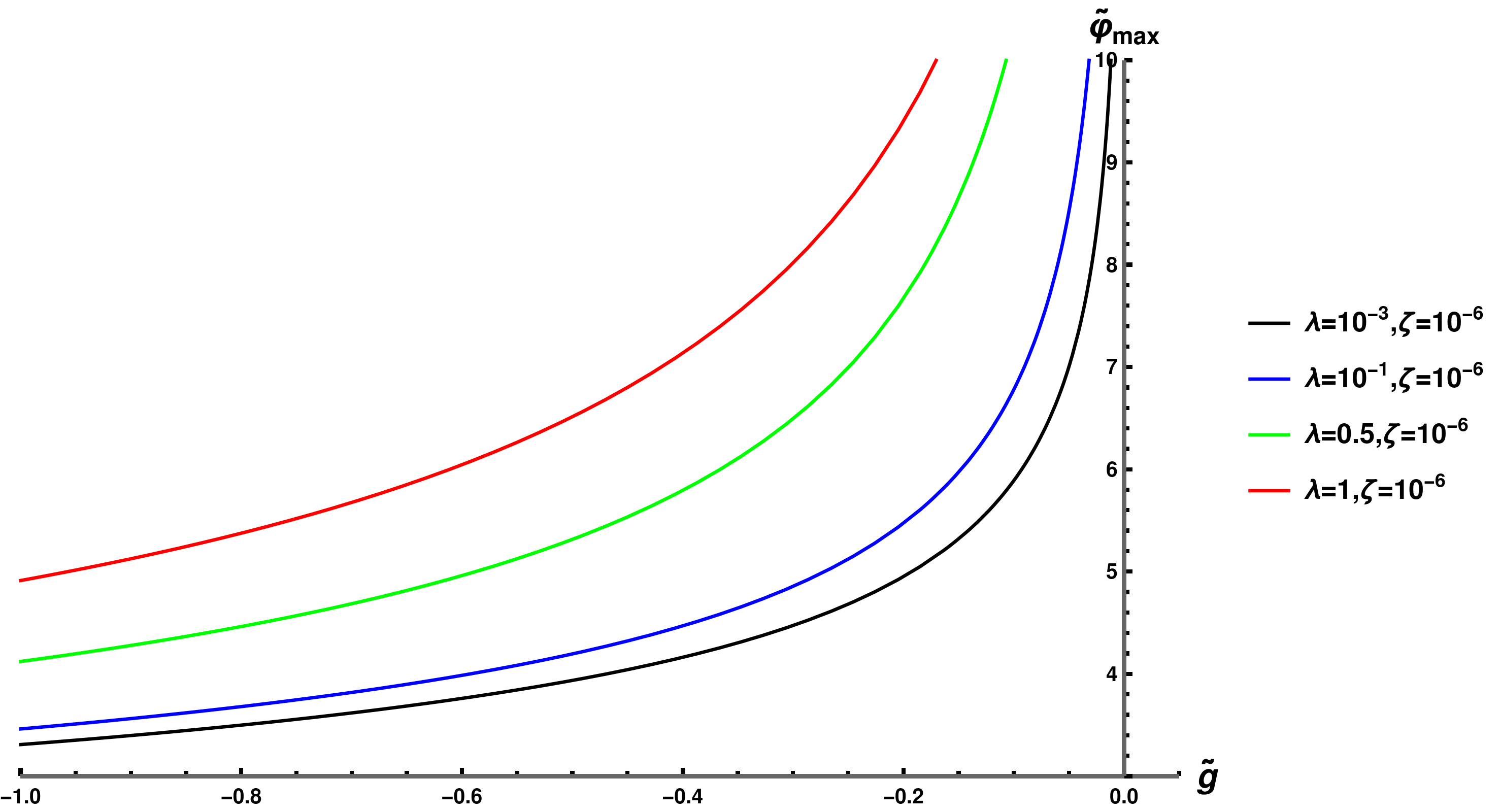}
  \end{center}
  \caption{The dependence of the position of the local maximum of effective potential \eqref{the_effective_potential_renormalized_minimally_reduced_variables} with respect to $\tilde{g}$}\label{plot_2}
\end{figure}

\begin{figure}[htb]
  \begin{center}
    \includegraphics[width=\textwidth]{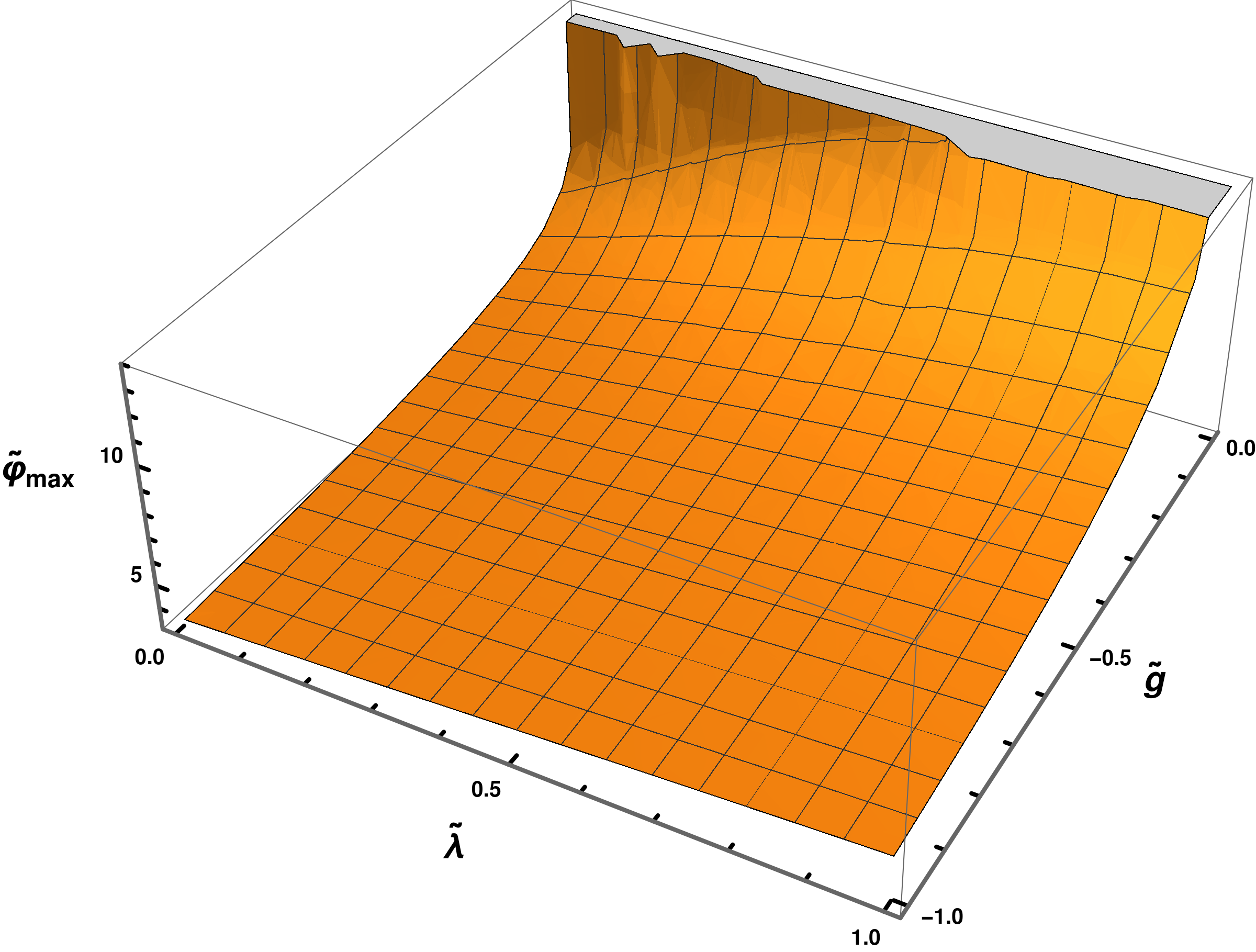}
  \end{center}
  \caption{The dependence of the position of the local maximum of effective potential \eqref{the_effective_potential_renormalized_minimally_reduced_variables} with respect to $\tilde{g}$ and $\lambda$. $\zeta=10^{-6}$.}\label{plot_3}
\end{figure}

Finally, if condition \eqref{criticality_condition_0} is met then the position of the local maximum also growth as the model approaches strong interaction regime ($\lambda\to 1$). We present plots showing such a behavior in Figure~\ref{plot_2} and Figure~\ref{plot_3}.

These results are important in the context of possible implications for inflation. As the 
potential~\eqref{the_effective_potential_renormalized_minimally} has an instability region, the local maximum marks 
the applicability region for the model. Initial conditions for inflation, in turn, shall also be taken from 
$\tilde\varphi\leq \tilde\varphi_\text{max}$. If condition~\eqref{criticality_condition_0} is fulfilled, 
then $\tilde\varphi_\text{max}$ can be calculated and direct constraints on the initial conditions compatible 
with the effective model can be established. If condition~\eqref{criticality_condition_0} is not fulfilled, 
then $\tilde\varphi_\text{max}$ is exponentially big and resides far beyond the Planck region. Therefore 
in that case no reasonable constraints on the initial conditions can be imposed.

\begin{figure}[htb]
  \begin{center}
    \begin{minipage}{\textwidth}
      \begin{minipage}{.49\textwidth}
        \includegraphics[width=\textwidth]{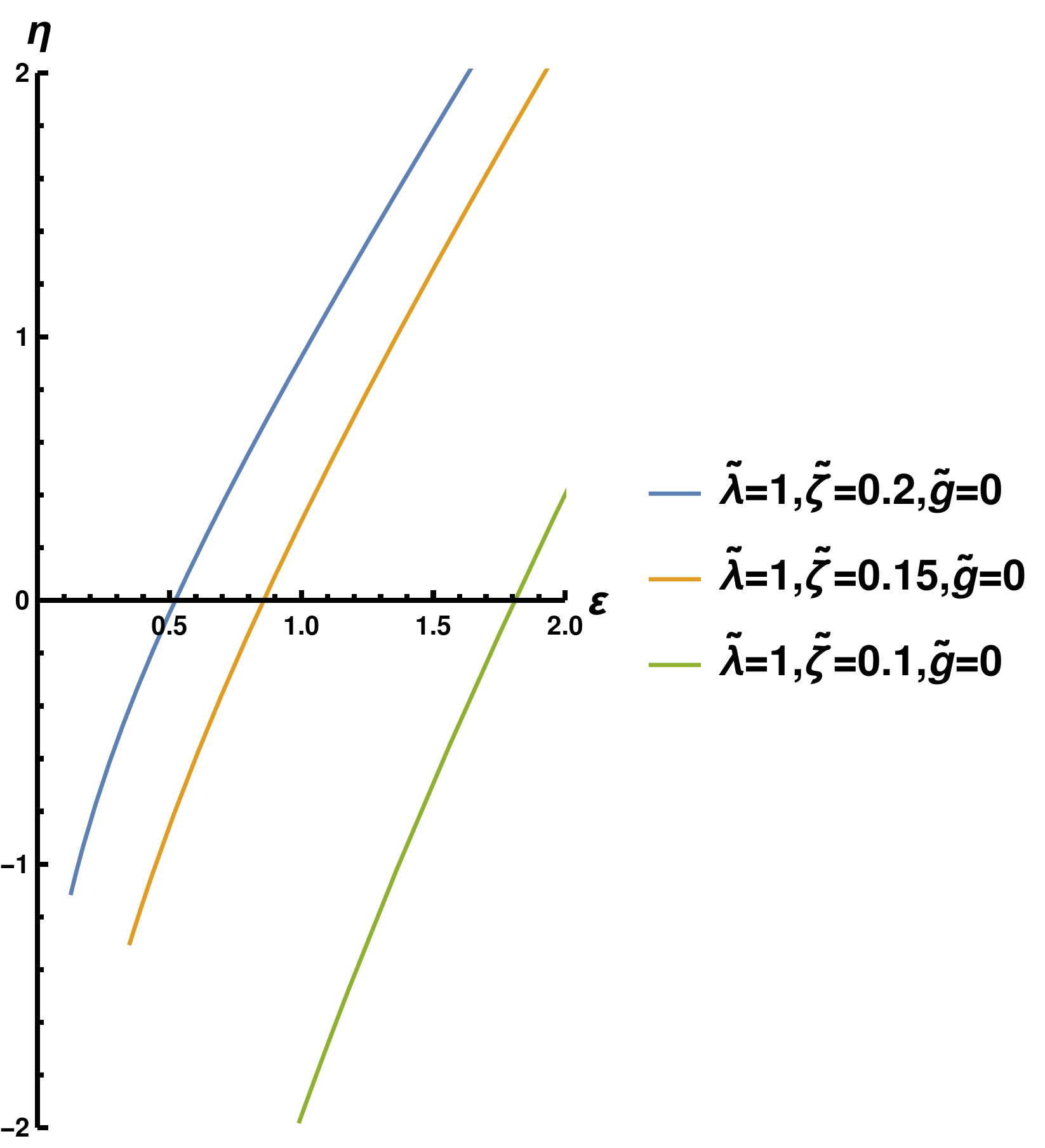}
      \end{minipage}
      \begin{minipage}{.49\textwidth}
        \includegraphics[width=\textwidth]{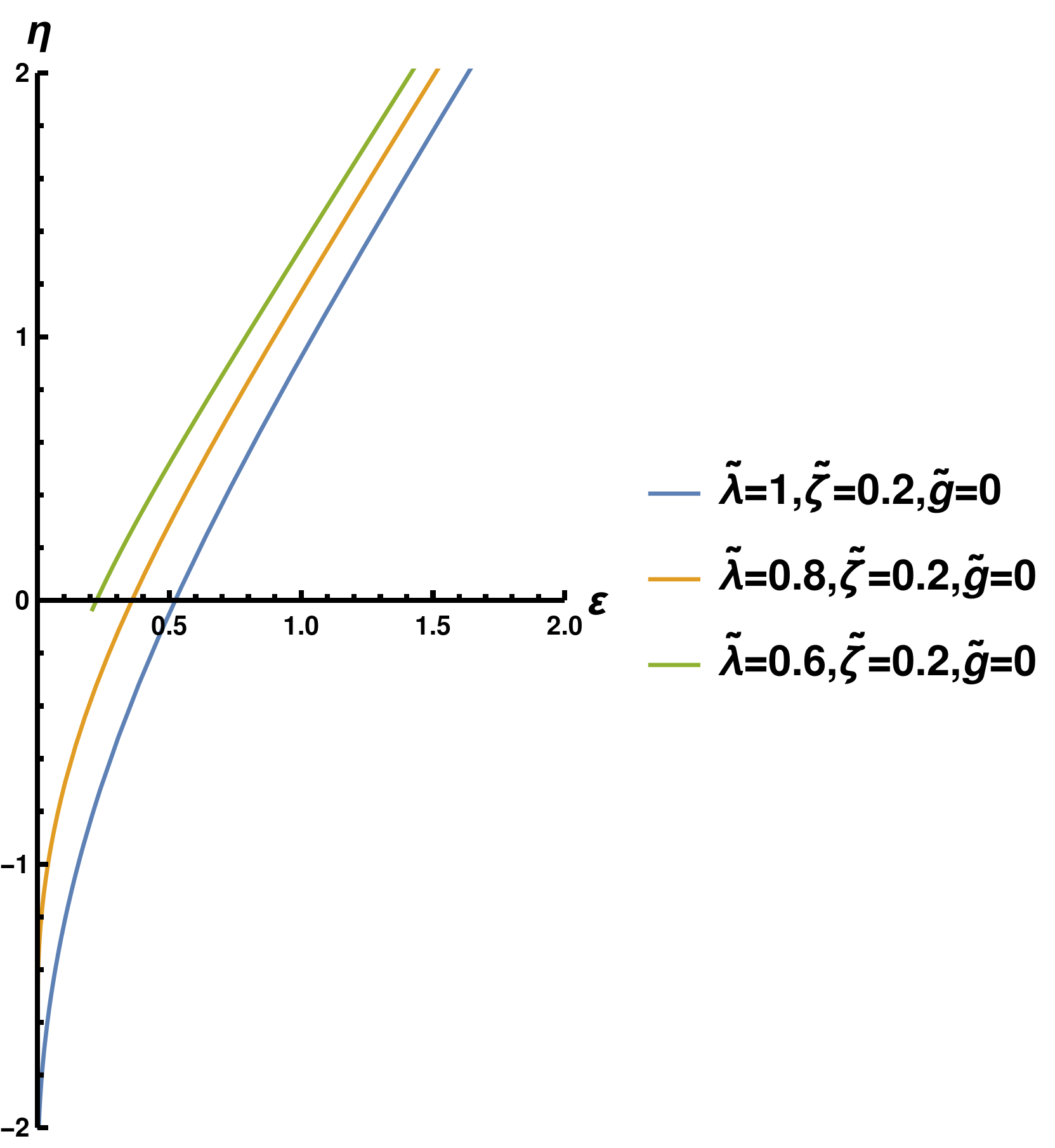}
      \end{minipage}
    \end{minipage}
  \end{center}

  \begin{center}
    \includegraphics[width=.49\textwidth]{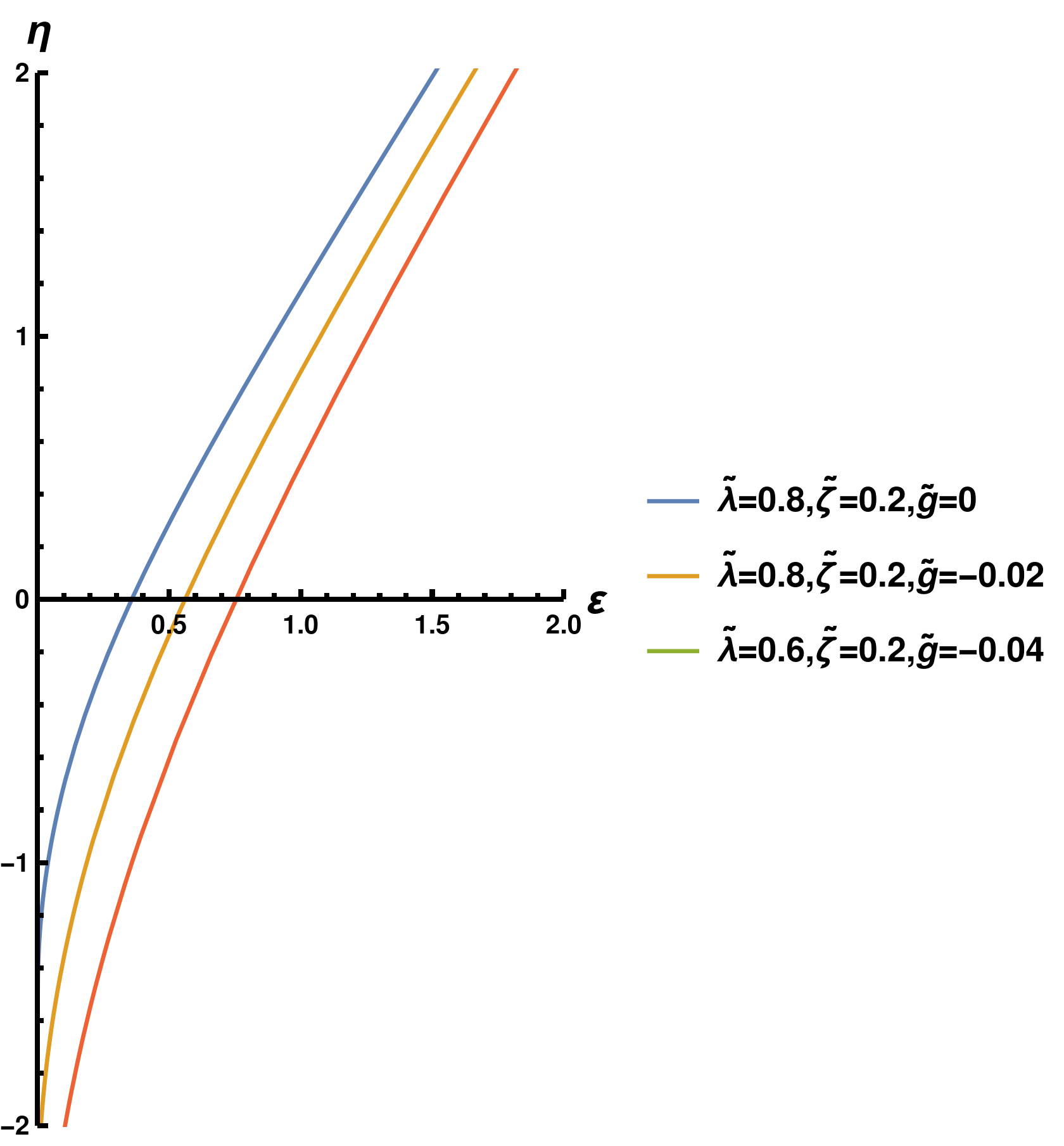}
  \end{center}
  
  \caption{Slow-roll parameters for the effective model. Lines start at infinity with $\tilde\varphi=0$ and strive to $\varepsilon=0$, $\eta=0$ region with $\tilde\varphi \to \tilde\varphi_\text{max}$.}\label{plot_4}
\end{figure}

The complete analysis of inflation scenarios described by the discussed effective model lies beyond the scope of this paper 
and will be presented elsewhere. However, simple numerical estimates provide a way to show that the model has 
at least one area of parameters where slow-roll conditions are satisfied. Let us focus on $\tilde{g}<0$ region. 
Slow-roll parameters evaluated via the scalar field effective potential~\eqref{the_effective_potential_renormalized_minimally} read
\begin{align}
  \varepsilon &= \cfrac23\,\cfrac{1}{\kappa^2} \,\left(\cfrac{1}{V_\text{eff,ren}} \, \cfrac{d}{d\varphi} \, V_\text{eff,ren}\right)^2\,, & \eta &=\cfrac43\,\cfrac{1}{\kappa^2}\,\cfrac{1}{V_\text{eff,ren}} \, \cfrac{d^2}{d\varphi^2}\,V_\text{eff,ren}\,.
\end{align}
In Figure \ref{plot_4} we present plots which show that for a certain range of parameters 
the effective potential enters an area where $\varepsilon <1$ and $\eta <1$. This area corresponds to 
$\tilde\varphi \sim \tilde\varphi_\text{max}$, therefore the models indeed has room for slow-roll inflationary scenarios.

\section{Discussion and conclusions}\label{section_discussion}

In this paper we studied a simple scalar-tensor model \eqref{the_model} and the effective potential generated by it. One-loop contributions to the effective potential up to $\mathcal{O}(\kappa^2)$ order is given by expression~\eqref{the_effective_potential}. 
The effective potential is regular in $\varphi\to 0$ limit therefore we choose the normalization point $\varphi_0=0$ and renormalize the effective potential with counter-term \eqref{the_counter-term}. The renormalized effective potential is given 
by~\eqref{the_effective_potential_renormalized_minimally}. It contains a new non-renormalizable effective 
$\varphi^6$ interaction dynamically generated at the one-loop level. 
The interaction is normalized at its observed value in the low-energy limit.

The effective potential \eqref{the_effective_potential_renormalized_minimally} always has an instability region. 
If condition~\eqref{criticality_condition_0} is satisfied, then the instability region exists in the weak field region. 
If condition~\eqref{criticality_condition_0} is not satisfied, then the instability region exists exponentially 
far in the strong field regime. The instability region naturally marks the effective model applicability domain. 
Finally, despite the fact that the potential is unstable, it does not generate new non-trivial minima in the stability region.

The main result of this paper is the renormalized effective potential \eqref{the_effective_potential_renormalized_minimally} 
while its applications shall be studied elsewhere. One of the main directions of further studies will be implementation 
of~\eqref{the_effective_potential_renormalized_minimally} for inflation. In Figure~\ref{plot_4} we present plots 
which explicitly show that it is possible to find a set of model parameters which guarantee the existence 
of a region with small slow-roll parameters. This fact shows that there is room for a standard slow-roll inflation 
within this model. A more detailed analysis of inflationary scenarios within this model will be performed in a forthcoming paper.

\section*{Acknowledgment}
The work (B.L.) was supported by the Foundation for the Advancement of Theoretical Physics and Mathematics “BASIS”.

\appendix

\section{Calculation of effective potentials}\label{section_factors}

In order to calculate an effective potential one shall calculate connected Green functions, recover vertex functions, and re-sum 
a series generating the effective potential. The technique is well described in~\cite{Buchbinder:1992rb} and we will not discus it in detail. However, the calculation of one-loop connected Green functions, which is the corner stone of the technique, must be discussed.

We use the continual integral technique to evaluate connected Green functions. An $n$-point one-loop connected Green function $G^c_n(x_1,\cdots,x_n)$ is given by the following expression \cite{Buchbinder:1992rb}:
\begin{align}\label{appendix_green_function_definition}
  \begin{split}
    i\, G^c_N(x_1,\cdots,x_N) = &\exp\left[i \, \mathcal{A}_\text{int} \left(\cfrac{\delta}{\delta(i J)} \,,\, \cfrac{\delta}{\delta(i J^{\mu\nu})}\right)\right] \\
    &\times \left(\prod\limits_{i=1}^n \cfrac{\delta}{\delta(i J(x_i))}\right) \left\{ \, \exp\left[\cfrac{i}{2} \, J\cdot \mathcal{G}\cdot J\right] \,\exp\left[\cfrac{i}{2} \, J^{\mu\nu}\cdot \mathcal{G}_{\mu\nu\alpha\beta}\cdot J^{\alpha\beta}\right]\right\} \Bigg|_{J,J^{\mu\nu}=0}^\text{connected}.
  \end{split}
\end{align}
Here $\mathcal{A}_\text{int} (\phi, h_{\mu\nu})$ is the part of action \eqref{the_model} describing interaction; $\mathcal{G}$ and $\mathcal{G}_{\mu\nu\alpha\beta}$ are Green functions
\begin{align}
  \mathcal{G}(x) &= \int\cfrac{d^4k}{(2\pi)^4} \, \cfrac{1}{-k^2 + m^2} \, e^{-i k x} \, , & \mathcal{G}_{\mu\nu\alpha\beta} (x) &= \cfrac12\, C_{\mu\nu\alpha\beta} \, \int\cfrac{d^4k}{(2\pi)^4} \, \cfrac{1}{-k^2} \, e^{-i k x}\,;
\end{align}
dots denote scalar products defined as follows:
\begin{align}
  J\cdot\mathcal{G}\cdot J &= \int d^4 x \, d^4 y \, J(x) \, \mathcal{G}(x-y) \, J(y)\,, & (\mathcal{G} \cdot J)(x) & = \int d^4 y \, \mathcal{G} (x-y)\, J(y)\,;
\end{align}
$J$, $J^{\mu\nu}$ are external currents. The notations on the right hand side of~\eqref{appendix_green_function_definition} note that we only account for contributions corresponding to connected diagrams with vanishing external currents. Expression \eqref{appendix_green_function_definition} can be simplified. Within the original expression~\eqref{appendix_green_function_definition} variational derivatives associated with an external field act on the exponent containing the scalar field Green function. Such an expression does contain contributions corresponding to both connected and disconnected diagrams. To exclude irrelevant contribution one shall only take the term that explicitly account for connected contributions which reads
\begin{align}\label{appendix_green_function_simplification}
  \begin{split} 
    i\, G^c_N(x_1,\cdots,x_N) = &\exp\left[i \, \mathcal{A}_\text{int} \left(\cfrac{\delta}{\delta(i J)} \, \cfrac{\delta}{\delta(i J^{\mu\nu})}\right)\right] \\
    &\times \left\{\left(\prod\limits_{i=1}^n (\mathcal{G} \cdot J)(x_i)\right) \, \exp\left[\cfrac{i}{2} \, J\cdot \mathcal{G}\cdot J\right] \,\exp\left[\cfrac{i}{2} \, J^{\mu\nu}\cdot \mathcal{G}_{\mu\nu\alpha\beta}\cdot J^{\alpha\beta}\right]\right\} \Bigg|_{J,J^{\mu\nu}=0}^\text{connected}.
  \end{split}
\end{align}
Here each variational derivative associated with an external line acted on each own Green function effectively spawning external lines. The remaining variation derivatives to be evaluated can only spawn internal lines or connections to external lines.

Within the discussed model at the leading order a $2N$-point connected Green functions read
\begin{align}
  &i \, G_{2N}^c(x_1,\cdots,x_{2N}) = \exp\left[-i\,\cfrac{\lambda}{4!} \int d^4 \zeta \, \left(\cfrac{\delta}{\delta(i J(\zeta))}\right)^4\right]\,\prod\limits_{i=1}^{2N}\cfrac{\delta}{\delta(i J(x_i))} \, \exp\left[\cfrac{i}{2}\, J\cdot\mathcal{G}\cdot J \right] \Bigg|_{J=0}^\text{connected}
\end{align}
\begin{align}
  \begin{split}
    &=\cfrac{1}{N!} \, \left(-i\,\cfrac{\lambda}{4!} \right)^N \begin{pmatrix} 4\\2,2\end{pmatrix}^N\,\int d^4 \zeta_1\cdots d^4 \zeta_N\left\{\left(\prod\limits_{i=1}^N\left(\cfrac{\delta}{\delta(i J(\zeta_i))}\right)^2\right)\left(\prod\limits_{j=1}^N (\mathcal{G}\cdot J)(x_j)\right)\right\}\\
      &\times \left\{\left(\prod\limits_{n=1}^N \left(\cfrac{\delta}{\delta(i J(\zeta_n))}\right)^2\right) \exp\left[\cfrac{i}{2} \, J \cdot \mathcal{G} \cdot J\right]\right\}\Bigg|_{J=0}^\text{connected} .
  \end{split}
\end{align}
Here factor $1/N!$ is due to the exponent expansion;
\begin{align}
  \begin{pmatrix} 4\\ 2,2 \end{pmatrix} = \cfrac{4!}{2! 2!}
\end{align}
is the multinomial coefficient that appears as we calculated derivatives of a product. The first term in curve brackets
\begin{align}
  \left(\prod\limits_{i=1}^N\left(\cfrac{\delta}{\delta(i J(\zeta_i))}\right)^2\right)\left(\prod\limits_{j=1}^N (\mathcal{G}\cdot J)(x_j)\right)
\end{align}
corresponds to external lines connected to internal interaction vertices. This contribution contains $(2N)!$ term which 
is the number of ways by which $2N$ external lines can be connected to an internal one-loop diagram. 
After an amputation of external lines all these contributions become equivalent so at the level of 
vertex functions contribution of this term is reduced to a $(2N)!$ factor. The second curve bracket term
\begin{align}
  \left\{\left(\prod\limits_{n=1}^N \left(\cfrac{\delta}{\delta(i J(\zeta_n))}\right)^2\right) \exp\left[\cfrac{i}{2} \, J \cdot \mathcal{G} \cdot J\right]\right\}\Bigg|_{J=0}^\text{connected}
\end{align}
describes scalar field propagators connecting internal interaction vertices. It contains many term but all of them 
are equivalent due to variable redefinition. After a suitable redefinition of variables, the contribution can be 
brought to the following form:
\begin{align}
  \left\{\left(\prod\limits_{n=1}^N \left(\cfrac{\delta}{\delta(i J(\zeta_n))}\right)^2\right) \exp\left[\cfrac{i}{2} \, J \cdot \mathcal{G} \cdot J\right]\right\}\Bigg|_{J=0}^\text{connected} = (N-1)!\, 2^{N-1} \, \cfrac{1}{i} \, \mathcal{G}(\zeta_1-\zeta_2) \cdots \cfrac{1}{i} \, \mathcal{G}(\zeta_N-\zeta_1).
\end{align}

Such a $2N$-point connected Green function generates the following contribution to the effective action $\Gamma$:
\begin{align}
  \begin{split}
    &\int d^4 \zeta_1 \cdots d^4 \zeta_{2N}\,\cfrac{1}{(2N)!}\, \Gamma_{2N}(\zeta_1,\cdots,\zeta_{2N})\,\varphi(\zeta_1) \cdots\varphi(\zeta_{2N})  \\
    &=\int d^4 x \, \cfrac{\varphi(x)^{2N}}{(2N)!} \,\cfrac{1}{N!} \, \left(-\cfrac{\lambda}{4!} \, \cfrac{4!}{2!}\right)^N\,\cfrac{1}{2^N} \, (2N)!\, (N-1)! \, 2^{N-1} \, (-i) \, \int\cfrac{d^4 k}{(2\pi)^4}\left( \cfrac{1}{-k^2+m^2}\right)^N
  \end{split}\\
  &=\int d^4 x (-i) \int\cfrac{d^4k}{(2\pi)^4}\, \cfrac{1}{2N} \, \left(\cfrac{\frac{\lambda}{2} \, \varphi^2}{k^2 -m ^2}\right)^N \, .
\end{align}
This expression (up to the inclusion of a non-vanishing mass term) matches the well-known result~\cite{Coleman:1973jx}.

Green functions that generate series \eqref{series_1} are defined similarly to the previous case:
\begin{align}
  \begin{split}
    &i\, G_{2N}^c\! =\! \cfrac{1}{2!} \int d^4 \!z_1 d^4\! z_2\left(-i\,\cfrac{\kappa}{4} \, m^2\right)^2 \eta^{\mu\nu} \eta^{\alpha\beta} \,\cfrac{\delta}{\delta(i J^{\mu\nu}(z_1))}\,\cfrac{\delta}{\delta(i J^{\alpha\beta}(z_2))} \left(\cfrac{\delta}{\delta(i J(z_1))}\right)^2 \left(\cfrac{\delta}{\delta(i J(z_2))}\right)^2\\
    &\times\cfrac{\left(-i\,\cfrac{\lambda}{4!}\right)^{N-1}}{(N-1)!}\int\left(\prod\limits_{n=1}^{N-1} d^4 \zeta_n\,\left(\cfrac{\delta}{\delta(i J(\zeta_n))}\right)^4 \right)\,\left\{\prod\limits_{i=1}^{2N}(\mathcal{G} \cdot J)(x_i)\,e^{\frac{i}{2}\,J\cdot\mathcal{G}\cdot J}\,e^{\frac{i}{2}\,J^{\mu\nu}\cdot\mathcal{G}_{\mu\nu\alpha\beta}\cdot J^{\alpha\beta}} \right\}\Bigg|_{J,J^{\mu\nu}=0}^\text{connected} \,.
  \end{split}
\end{align}
Scalar quartic interaction and scalar-graviton interaction can be placed in two independent exponents as $\delta/\delta(i J)$ commutes with $\delta/\delta(i J^{\mu\nu})$. Factor $1/2!$ is due to an expansion of the exponent describing gravitational interaction. Factor $1/(N-1)!$ is due to an expansion of scalar field interaction exponent. This expression can be simplified in a similar way:
\begin{align}
  \begin{split}
    i\, G_{2N}^c =& \cfrac{1}{2!} \int d^4 \!z_1 d^4\! z_2 \left(-i\,\cfrac{\kappa}{4} \, m^2\right)^2\,\eta^{\mu\nu} \, \eta^{\alpha\beta} \,\cfrac{1}{i}\,\mathcal{G}_{\mu\nu\alpha\beta}(z_1-z_2) \left(\cfrac{\delta}{\delta(i J(z_1))}\right)^2 \left(\cfrac{\delta}{\delta(i J(z_2))}\right)^2\\
    &\times\cfrac{\left(-i\cfrac{\lambda}{4!}\right)^{N-1}}{(N-1)!}\int\left(\prod\limits_{n=1}^{N-1} d^4 \zeta_n\,\left(\cfrac{\delta}{\delta(i J(\zeta_n))}\right)^4 \right)\,\left\{\prod\limits_{i=1}^{2N}(\mathcal{G} \cdot J)(x_i)\,\exp\left[\cfrac{i}{2}\,J\cdot\mathcal{G}\cdot J \right] \right\}\Bigg|_{J=0}^\text{connected} \,.
  \end{split}
\end{align}
\begin{align}
  \begin{split}
    =&\cfrac{1}{2!} \int d^4\! z_1 \, d^4\! z_2\,\left(-i\,\cfrac{\kappa}{4} \, m^2\right)^2\,\cfrac{\left(-i\cfrac{\lambda}{4!}\right)^{N-1}}{(N-1)!}\begin{pmatrix} 2\\1,1\end{pmatrix}^2\! \begin{pmatrix}4\\2,2\end{pmatrix}^{N-1} \, \int \left(\prod\limits_{n=1}^{N-1} d^4\!\zeta_n \right)\\
        &\times\left\{ \prod\limits_{i=1}^2\,\cfrac{\delta}{\delta(i J(z_i))}\,\prod\limits_{i=1}^{N-1}\left(\cfrac{\delta}{\delta(i J(\zeta_i))}\right)^2\,\left(\prod\limits_{i=1}^{2N}(\mathcal{G} \cdot J)(x_i)\right)\right\}\\
        &\times\left\{\eta^{\mu\nu} \, \eta^{\alpha\beta} \,\cfrac{1}{i}\,\mathcal{G}_{\mu\nu\alpha\beta}(z_1-z_2)\prod\limits_{i=1}^2\,\cfrac{\delta}{\delta(i J(z_i))}\,\prod\limits_{i=1}^{N-1}\left(\cfrac{\delta}{\delta(i J(\zeta_i))}\right)^2\,\times\exp\left[\cfrac{i}{2}\,J\cdot\mathcal{G}\cdot J \right]\right\} \Bigg|_{J=0}^\text{connected} \,.
  \end{split}
\end{align}
At the level of the effective action this contribution can be simplified in exactly the same way as the LO contribution. The first expression in curved brackets describes all possible ways to connect $2N$ external lines to an internal structure. At the level of the effective action this contribution is reduced to $(2N)!$ factor. The second expression in curved brackets describes all ways to arrange internal lines. In full analogy with the previous case all term in this expression match up to a redefinition of integration variable. The corresponding contribution will generate factor $(N-1)! 2^{N-1}$ in the effective action.The corresponding part of the effective action reads:
\begin{align}
  \begin{split}
    &\int d^4 \zeta_1 \cdots d^4 \zeta_{2N}\,\cfrac{1}{(2N)!}\, \Gamma_{2N}(\zeta_1,\cdots,\zeta_{2N})\,\varphi(\zeta_1) \cdots\varphi(\zeta_{2N})  \\
    &=\int d^4 x (-i) \cfrac{\kappa^2 m^4}{\lambda} \int \cfrac{d^4 k}{(2\pi)^4}\,\cfrac{1}{-k^2} \,\left(\cfrac{\frac{\lambda}{2} \, \varphi^2}{k^2-m^2}\right)^N \,.
  \end{split}
\end{align}

Similar calculations are performed for series \eqref{series_2}. Green functions are given by the following expression (with $N\geq 2$):
\begin{align}
  \begin{split}
    i G_{2N}^c=&\int d^4 \!z_1\! \left(-i\cfrac{\kappa}{2}\,\cfrac{\lambda}{4!}\right) \eta^{\mu\nu} \cfrac{\delta}{\delta(i J^{\mu\nu}(z_1))}\left(\cfrac{\delta}{\delta(i J(z_1))}\right)^4 \int d^4\! z_2 \left(-i\cfrac{\kappa}{4} \right) \, m^2 \eta^{\alpha\beta} \cfrac{\delta}{\delta(i J^{\alpha\beta}(z_2))}\left(\cfrac{\delta}{\delta(i J(z_2))}\right)^2\\
    &\int\left(\prod\limits_{i=1}^{N-2} d^4 \zeta_i \left(\cfrac{\delta}{\delta(i J(\zeta_i))}\right)^4\right) \left(-i\cfrac{\lambda}{4!}\right)^{N-2} \, \cfrac{1}{(N-2)!}\left[\prod\limits_{j=1}^{2N} (\mathcal{G}\cdot J)(x_j) \,e^{\frac{i}{2}\,J\cdot\mathcal{G}\cdot J}\,e^{\frac{i}{2}\,J^{\mu\nu}\cdot\mathcal{G}_{\mu\nu\alpha\beta}\cdot J^{\alpha\beta}} \right]
  \end{split}
\end{align}
The same considerations allow one to bring the expression to the following form:
\begin{align}
  \begin{split}
    i G_{2N}^c =&\left(-i\cfrac{\kappa}{2}\, \cfrac{\lambda}{4!}\right)\left(-i\cfrac{\kappa}{4}\right)\,m^2 \,\cfrac{1}{(N-2)!} \,\left(-i\cfrac{\lambda}{4!}\right)^{N-2} \begin{pmatrix} 4\\3,1\end{pmatrix} \begin{pmatrix} 2\\1,1\end{pmatrix} \begin{pmatrix} 4\\2,2\end{pmatrix}^{N-2}\int d^4\! z_1 d^4\! z_2 \prod\limits_{n=1}^{N-2} d^4\! \zeta_n \\
          &\times\eta^{\mu\nu}\cfrac{1}{i}\,\mathcal{G}_{\mu\nu\alpha\beta}(z_1-z_2)\eta^{\alpha\beta}\, \left\{ \left(\cfrac{\delta}{\delta(i J(z_1))}\right)^3 \cfrac{\delta}{\delta(i J(z_2))}\prod\limits_{i=1}^{N-2} \left(\cfrac{\delta}{\delta(i J(\zeta_i))}\right)^2 \prod\limits_{j=1}^{2N} (\mathcal{G}\cdot J)(x_j)\right\}\\
          &\times \left\{\eta^{\mu\nu}\,\cfrac{1}{i}\, \mathcal{G}_{\mu\nu\alpha\beta}(z_1-z_2) \eta^{\alpha\beta}\cfrac{\delta}{\delta(i J(z_1))}\cfrac{\delta}{\delta(i J(z_2))}\prod\limits_{l=1}^{N_2} \left(\cfrac{\delta}{\delta(i J(\zeta_l))}\right)^2 \exp\left[\cfrac{i}{2} \, J\cdot\mathcal{G}\cdot J\right]\right\} \Bigg|_{J=0}^\text{connected}.
  \end{split}
\end{align}
In full analogy with the previous cases the first term in curved brackets describes external scalar lines and at the level of an effective action reduces to a factor $(2N)!$. The second term describes the structure of internal lines connecting interaction vertices and at the level of effective action it is reduced to factor $(N-2)! 2^{N-2}$. The factor depends on $N-2$ as the Green function $G_{2N}^c$ contains only $N-2$ quartic scalar interaction vertices. Finally, such a Green function generates the following contribution to the effective action:
\begin{align}
  \begin{split}
    &\int d^4 \zeta_1 \cdots d^4 \zeta_{2N}\,\cfrac{1}{(2N)!}\, \Gamma_{2N}(\zeta_1,\cdots,\zeta_{2N})\,\varphi(\zeta_1) \cdots\varphi(\zeta_{2N})\\
    &=\int d^4 x \,(-i) \int\cfrac{d^4 k}{(2\pi)^4} \,\cfrac{\kappa^2 m^2}{3} \, \varphi^2 \,\cfrac{1}{-k^2} \left(\cfrac{\frac{\lambda}{2} \, \varphi^2}{k^2 -m^2}\right)^{N-1}.
  \end{split}
\end{align}

Finally, the same actions shall be done with series \eqref{series_3}. The green function generating the series read ($N\geq 3$):
\begin{align}
  \begin{split}
    i\,G_{2N} =&\cfrac12\,\int d^4\!z_2 d^4 \!z_2\left(-i\,\cfrac{\kappa}{2}\,\cfrac{\lambda}{4!}\right)^2\,\eta^{\mu\nu}\eta^{\alpha\beta}\,\cfrac{\delta}{\delta(i J^{\mu\nu}(z_1))}\cfrac{\delta}{\delta(i J^{\alpha\beta}(z_2))} \left(\cfrac{\delta}{\delta(i J(z_1))}\right)^4\left(\cfrac{\delta}{\delta(i J(z_2))}\right)^4 \\
    &\cfrac{1}{(N-3)!}\,\left(-i\cfrac{\lambda}{4!}\right)^{N-3}\int \left(\prod\limits_{i=1}^{N-3}d^4\!\zeta_i\left(\cfrac{\delta}{\delta(i J(\zeta_i))}\right)^4\right)\left[\prod\limits_{j=1}^{2N} \,(\mathcal{G}\cdot J)(x_j)\,e^{\frac{i}{2}\,J\cdot\mathcal{G}\cdot J}\,e^{\frac{i}{2}\,J^{\mu\nu}\cdot\mathcal{G}_{\mu\nu\alpha\beta}\cdot J^{\alpha\beta}} \right]
  \end{split}
\end{align}
\begin{align}
  \begin{split}
    =& \cfrac12\left(-i\,\cfrac{\lambda}{4!}\right)^{N-1}\left(\cfrac{\kappa}{2}\right)^2\,\cfrac{1}{(N-3)!} \begin{pmatrix}4\\3,1\end{pmatrix}^2 \begin{pmatrix}4\\2,2\end{pmatrix}^{N-3} \int d^4\!z_1 d^4\!z_2 \prod\limits_{i=1}^{N-3} d^4\zeta_i \\
        &\times \eta^{\mu\nu}\cfrac{1}{i}\,\mathcal{G}_{\mu\nu\alpha\beta}(z_1-z_2)\eta^{\alpha\beta} \left\{ \left(\cfrac{\delta}{\delta(i J(z_1))}\right)^3 \left(\cfrac{\delta}{\delta(i J(z_2))}\right)^3 \prod\limits_{i=1}^{N-3} \left(\cfrac{\delta}{\delta(i J(\zeta_i))}\right)^2 \prod\limits_{j=1}^{2N} (\mathcal{G}\cdot J)(x_j) \right\}\\
        & \left\{ \cfrac{\delta}{\delta(i J(z_1))}\cfrac{\delta}{\delta(i J(z_2))}\prod\limits_{i=1}^{N-3} \left(\cfrac{\delta}{\delta(i J(\zeta_i))}\right)^2 \, \exp\left[\cfrac{i}{2} \, J \cdot\mathcal{G}\cdot J\right] \right\}\,.
  \end{split}
\end{align}
This expression, in turn, provides the following contribution to the effective action:
\begin{align}
  \begin{split}
    &\int d^4 \zeta_1 \cdots d^4 \zeta_{2N}\,\cfrac{1}{(2N)!}\, \Gamma_{2N}(\zeta_1,\cdots,\zeta_{2N})\,\varphi(\zeta_1) \cdots\varphi(\zeta_{2N})\\
    &=\int d^4 x \, (-i)\,\int\cfrac{d^4k}{(2\pi)^4} \,\cfrac{\kappa^2 \lambda\,\varphi^4}{36}  \,\cfrac{1}{-k^2} \, \left(\cfrac{\frac{\lambda}{2}\,\varphi^2}{k^2-m^2}\right)^{N-2}\,.
  \end{split}
\end{align}

\bibliographystyle{unsrturl}
\bibliography{SFEPN.bib}

\newpage

\newpage

\newpage

\newpage

\end{document}